\documentclass[%
 reprint,
superscriptaddress,
 amsmath,amssymb,
 aps,
]{revtex4-1}

\usepackage{graphicx}

\usepackage{dcolumn}
\usepackage{bm}
\usepackage{xcolor}

\begin{document}


\title{Sensitivity for four-body tau-lepton decays at Belle and Belle II experiments}

\author{I. Dom\'inguez}
\author{P. S. Mares Damas}
\author{P. L. M. Podesta Lerma}

\affiliation{Facultad de Ciencias F\'isico-Matem\'aticas, Universidad Aut\'onoma de Sinaloa, Avenida de las Am\'ericas y Boulevard Universitarios, Ciudad Universitaria, C.P. 80000, Culiac\'an, Sinaloa, M\'exico.}

 \author{D. Rodr\'iguez P\'erez}
\affiliation{Facultad de Informática Culiacán, Universidad Aut\'onoma de Sinaloa, Josefa Ort\'iz de Dominguez S/N, Ciudad Universitaria, C.P. 80013, Culiac\'an, Sinaloa, M\'exico.}

\date{\today}

\begin{abstract}
  We study the expected sensitivity at Belle and Belle II for four-body $\tau^\mp \to X^\pm l^\mp l^\mp \nu_{\tau}$ decays where $l=e$
  or $\mu$ and $X=\pi$, $K$, $\rho$ and $K^*$ mesons.  These decay processes violate the total lepton number ($|\Delta L|=2$ ) and they can be induced by the exchange of Majorana neutrinos. In particular, we consider  lifetimes in the accessible ranges of $\tau_N$ = 5, 100 ps and extract the limits on $|V_{\ell N}|^2$ without any additional assumption on the relative size of the mixing matrix elements. For an integrated luminosity collected of 1 ab$^{-1}$ at Belle, we found significant sensitivity on branching fractions of the order BR($\tau^\mp \to X^\pm l^\mp l^\mp \nu_{\tau}$) $\sim 10^{-8}$. For an integrated luminosity expected of 50 ab$^{-1}$ and intermediate luminosity of 10 ab$^{-1}$ at the Belle II, we found significant sensitivity on branching fractions of the order BR($\tau^\mp \to X^\pm l^\mp l^\mp \nu_{\tau}$) $\sim 10^{-9}-10^{-8}$. We use these sensitivities to set limits for the exclusion regions on the parameter space $(m_N, |V_{\ell N}|^2)$ associated with the heavy neutrino; such that for a $|V_{\ell N}|^2 \sim \mathcal{O}(10^{-5})$ at $\tau_N = 100$ ps, we find the bounds as $0.140 < m_N < 1.776$ GeV for $\tau^- \to X^+ e^- e^- \nu_\tau$ and $0.245 < m_N < 1.671 $ GeV for $\tau^- \to X^+ \mu^- \mu^- \nu_\tau$.
  
\end{abstract}

\maketitle


\section{\label{sec1}Introduction}

The neutrino oscillation experiments \citep{Fukuda:1998mi, Wendell:2010md, Ambrosio:2003yz} have given indications of physics beyond the Standard Model (SM), one of them is the fact that neutrinos should be particles with nonzero masses \citep{Strumia:2006db}. At present, the SM only incorporates left-handed neutrinos $\nu_L$ in $SU(2)_L$ gauge group doubles and as consequence the neutrino mass term cannot be constructed. To include the neutrino mass term in the SM different extensions have been proposed, some of them consider: left-right symmetric gauge theories \citep{PhysRevD.10.275}, models with exotic Higgs representations \citep{ZEE1980389} or  R-parity violation interactions ($\Delta L = 1$) in Supersymmetry with extra dimensions \citep{AULAKH1982136,Hall:1983id}.
In this paper, we focus in the Seesaw Type I mechanism \citep{MINKOWSKI1977421} which introduce $n$ right-handed SM singlet neutrino fields $N_R$. These fields allowed to couple to their own charge conjugate field to form Majorana mass terms. This mechanism implies the  lepton number violation (LNV), which is one of main feature respect to other models \citep{PhysRevD.10.275,ZEE1980389,AULAKH1982136,Hall:1983id}.

The decays in the SM conserve the lepton number, then a signal of LNV is a probe of new physics (NP). Previous studies in Seesaw Type I mechanism consider the number of right-handed singlet neutrino fields as $n = 1$ \citep{Atre:2009rg,Helo:2010cw} to estimate the three-body lepton decays $\tau^- \to X^+ \ell^- \nu_\tau$ where $X= K$, $D$ and $B$ and $\ell = e$, $\mu$. Furthermore, recent studies also consider four-body decays with an enhancement in the branching fractions \citep{Quintero:2011yh, Castro:2012ma, Castro:2012gi}.  

In this work, we study sensitivity four-body $\tau$ lepton decays  $\tau^- \to X^+ \ell^- \ell^- \nu_\tau$ with $\ell = \mu$ or $e$ and $X=\pi$, $K$, $\rho$ or $K^*$ and their conjugate decays, at Belle II energies \citep{Abe:2010gxa}. These decays have been previously study in \citep{Castro:2012gi} considering the decay width of the heavy Majorana neutrino as the contribution of low energy tau decays and rare meson decays, with $m_N \ll m_W$.  However, we consider the heavy neutrino lifetime of $\tau_N$ = 5, 100 ps, where the detector has the similar sensitivity as the decays considered in Sec. \ref{sec3}. This allows us to extract the limits on $|V_{\ell N}|^2$ without any additional assumption on the relative size of the mixing matrix elements. Based in the sensitivity of Belle II we can constrain the parameter  space ($m_N$, $\vert V_{\ell N}\vert^2$). 

This paper is organized as follow. In Sec. \ref{sec2} we study the four-body LNV decays of the $\tau$ lepton. The Sec. \ref{sec3} contains the experimental sensitivity for these channels at Belle and Belle II experiments. In Sec. \ref{sec4} we present the exclusion regions on the parameter space $(m_N, |V_{\ell N}|^2)$ of the heavy neutrino that can be achieved from the experimental searched at Belle and Belle II. Finally, our conclusion are given in Sec. \ref{sec5}.

\section{\label{sec2}Four-body $|\Delta L| = 2$ decays of $\tau$ lepton}

\begin{figure}[h]
\begin{center}
  \includegraphics[width=1\linewidth]{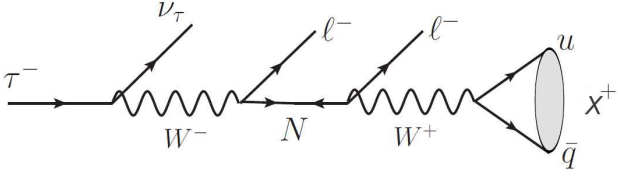}
\end{center}
\caption{\label{fig:diagram} Feynman diagram to LNV in four-body tau decays $\tau^- \to X^+ \ell^- \ell^- \nu_\tau$ ($X=\pi$, $K$, $\rho$ or $K^*$) induced by a Majorana neutrino N.}
\end{figure}

We consider the four-body $|\Delta L| = 2$ decays of the $\tau$ lepton $\tau^- \to  X^+ \ell^- \ell^- \nu_\tau$ (Fig. \ref{fig:diagram}), with $\ell = \mu$ or $e$ and $X=\pi$, $K$, $\rho$ or $K^*$. In the framework of Seesaw Type I mechanism and the assumption that only one heavy Majorana neutrino $N$, these processes occur via the intermediate on-shell Majorana neutrino through the leptonic decay $\tau^- \to \nu_\tau \ell^- N$ followed by the subsequent semileptonic decay $N \to \ell^- X^+$, with a kinematically allowed mass range for the corresponding channel, see Table \ref{table1}.

\begin{table}[h]
\caption{\label{table1} Four-body $|\Delta L| = 2$ decays of the $\tau$ lepton.}
\begin{ruledtabular}
\begin{tabular}{lc}
Decay mode & $m_N$ [GeV] \\
\colrule
$\tau^- \to \pi^+ e^- e^- \nu_\tau$ & 0.140 - 1.776 \\
$\tau^- \to K^+ e^- e^- \nu_\tau$ & 0.494 - 1.776 \\
$\tau^- \to \rho^+ e^- e^- \nu_\tau$ & 0.776 - 1.776 \\
$\tau^- \to K^{*+} e^- e^- \nu_\tau$ & 0.892 - 1.776 \\
$\tau^- \to \pi^+ \mu^-\mu^- \nu_\tau$ & 0.245 - 1.671 \\
$\tau^- \to K^+ \mu^-\mu^- \nu_\tau$ & 0.599 - 1.671 \\
$\tau^- \to \rho^+ \mu^-\mu^- \nu_\tau$ & 0.861 - 1.671 \\
$\tau^- \to K^{*+} \mu^-\mu^- \nu_\tau$ & 0.997 - 1.671 \\
\end{tabular}
\end{ruledtabular}
\end{table}

Then, the decays are splitted into two subprocesses and the corresponding branching fraction can be written in the factorized form 
\begin{eqnarray}
  BR(\tau^- \to X^+ \ell^-\ell^- \nu_\tau) & = &  BR(\tau^- \to \nu_\tau \ell^- N) \nonumber\\
   & & \times \Gamma(N \to \ell^- X^+)\tau_N / \hslash
   \label{eq:factorForm}
\end{eqnarray}
with $\tau_N$ as the timelife of the Majorana neutrino. To obtain the branching fraction of the leptonic subprocess $\tau^- \to \nu_\tau \ell^- N$, we begin from its amplitude which is given by the expression 
\begin{eqnarray}
  \mathcal{M}(\tau^- \to \ell^- N \nu_\tau) & = & -i\frac{G_F}{\sqrt{2}} V_{\ell N}
  [\bar{\nu}_\tau(p_\nu) \gamma^\mu (1-\gamma^5) \tau(p_\tau) ]\nonumber\\
    & & \times[\bar{\ell}(p_\ell) \gamma^\mu (1-\gamma^5) N(p_N)]
 \label{eq:wideeq2}
\end{eqnarray}
where $G_F$ is the Fermi constant and $V_{\ell N}$ is the coupling of the heavy neutrino N to the charged current of lepton flavor $\ell$. Furthermore, using the three-body decay kinematics we compute the decay width which is given as

\begin{equation}
d^2\Gamma =\frac{1}{4} \frac{1}{4(4\pi)^3 m_\tau^3} |\mathcal{M}|^2 ds_{12}ds_{13}
\end{equation}
where the global factor 1/4 implies the identical final state leptons (1/2) and the spin average of the initial particle (1/2); the region of integration for $s_{12} = (p_\nu + p_\ell)^2$ and $s_{13} = (p_\nu + p_N)^2 $ are   

\begin{equation}
  s_{13}^- \leqslant s_{13} \leqslant s_{13}^+, \hspace{.5cm} (m_\nu + m_\ell)^2 \leqslant s_{12} \leqslant (m_\tau - m_N)^2,
\end{equation}
where we used the K\"{a}llen function which is denoted by $\lambda(x,y,z) = x^2 + y^2 + z^2 - 2(xy+xz+yz)$ to write the limits of integration of $s_{13}^\pm$ as 

\begin{eqnarray}
 s_{13}^\pm  &  =  &  \frac{1}{2s_{12}} \Bigg \lbrace  2(m_\tau^2 + m_\nu^2 - s_{12})s_{12} \nonumber \\
 &   &- (m_\tau^2 - m_N^2-s_{12}) (s_{12} - m_\nu^2 + m_\ell^2)\nonumber\\
  &  &  \pm \sqrt{\lambda(m_\tau^2, m_N^2,s_{12})(\lambda(s_{12}, m_\nu^2,m_\ell^2)) } \Bigg \rbrace.
 \label{eq:wideeq3}
\end{eqnarray}

The corresponding momentum scalar product are expressed in the following set of Mandelstam variables

\begin{eqnarray}
  p_\tau \cdot p_\nu & = &  (s_{12} + s_{13} - m_\ell^2 - m_N^2 )/ 2, \nonumber\\
  p_\tau \cdot p_\ell & = &  (m_\tau^2 + m_\ell^2 -s_{13} )/ 2, \nonumber\\
  p_\tau \cdot p_N & = &  (m_\tau^2 + m_N^2 -s_{12} )/ 2, \nonumber\\
  p_\nu \cdot p_\ell & = & (s_{12} - m_\nu^2 - m_\ell^2)/ 2, \nonumber\\   
  p_\nu \cdot p_N & = & (s_{13} - m_\nu^2 - m_N^2)/ 2, \nonumber\\   
  p_\ell \cdot p_N & = & (m_\tau^2 + m_\nu^2 - s_{12} - s_{13})/ 2.
\end{eqnarray} 

Thereby, the decay width for $\tau \to \nu \ell N$ can be written as 
\begin{widetext}
\begin{equation}
  \Gamma(\tau \to \nu_\tau \ell N) =\frac{1}{4} \frac{G_F^2}{64m_\tau^3(2\pi)^3} |V_{\ell N}|^2 \int_{s_{12}^-}^{s_{12}^+}\int_{s_{13}^-}^{s_{13}^+}  64(s_{12}-m_\ell^2 - m_\nu^2 )(m_N^2+m_\tau^2-s_{12})\, ds_{13}\, ds_{12},
  \label{eq:wideeq4}
\end{equation}
\end{widetext}
and the $BR(\tau^- \to \nu_\tau \ell^- N)$ is then obtained dividing  (\ref{eq:wideeq4}) by the total decay width of $\tau$ lepton, taken from \citep{PhysRevD.98.030001}.

On the other hand, the decay width of $\Gamma(N \to \ell^- X^+)$ is given by the expression

\begin{eqnarray}
   \mathcal{M}(N \to \ell^- X^+) & = & -i \frac{G_F}{\sqrt{2}} V_{uq}^{CKM} V_{\ell N}\nonumber\\
  & & \times \left[ \bar{\ell}(p_\ell) \gamma_\mu (1-\gamma^5)  N(p_N) \right] X^\mu
 \label{eq:wideeq2}
\end{eqnarray} 
where $V_{uq}^{CKM}$ (with $q=d$ for $\pi$ and $\rho$; and $q=s$ for $K$ and $K^*$) is the up-down Cabibbo-Kobayashi-Maskawa (CKM) matrix element associated to the hadronic vertex $X^\mu$, which the corresponding term to the pseudoscalar and vector mesons is 

\begin{equation}
X^\mu = \begin{cases} -if_X p_X^\mu, & X = \pi, K \\ f_X m_X \epsilon_X^\mu, & X = \rho, K^* \end{cases},
\end{equation}
where $\epsilon_X^\mu$ is the polarized vector meson. We compute the decay width which is given by the two-body decay kinematics, defined as

\begin{equation}
d\Gamma = \frac{1}{2} \frac{1}{32\pi^2} |\mathcal{M}|^2 \frac{|p_\ell|}{m_\tau}d\Omega
\label{width2}
\end{equation}
where $d\Omega$ is the differential solid angle. The corresponding momentum scalar product are expressed in the following form

\begin{eqnarray}
  p_N \cdot p_\ell & = & (m_N^2 + m_\ell^2 - m_X^2)/ 2, \nonumber\\ 
  p_N \cdot p_X & = & (m_N^2 + m_X^2 - m_\ell^2)/ 2, \nonumber\\   
  p_\ell \cdot p_X & = & (m_N^2 - m_\ell^2 - m_X^2)/ 2.
\end{eqnarray} 

The total decay width is then obtained by computing the square amplitude (\ref{eq:wideeq2}) and performing the integral (\ref{width2}). Then, the result for pseudoscalar mesons $(X = \pi, K)$ is 

\begin{eqnarray}
\Gamma(N \to \ell^- X^+) & = & \frac{G_F^2}{16\pi}|V_{uq}^{CKM}|^2 |V_{\ell N}|^2 f_X^2  m_N \nonumber\\
 &  & \times \sqrt{\lambda(m_N^2,m_\ell^2,m_X^2)} \Bigg [ \left(1- \frac{m_\ell^2}{m_N^2} \right)^2 \nonumber\\
 &  &- \frac{m_X^2}{m_N^2} \left(1+ \frac{m_\ell^2}{m_N^2} \right) \Bigg ],\label{dwidth21}
\end{eqnarray}
by the other hand, for vector mesons $(X = \rho, K^*)$ we have 
\begin{eqnarray}
\Gamma(N \to \ell^- X^+) & = & \frac{G_F^2}{16\pi}|V_{uq}^{CKM}|^2 |V_{\ell N}|^2 f_X^2  m_N \nonumber\\
 &  & \times \sqrt{\lambda(m_N^2,m_\ell^2,m_X^2)} \Bigg [ \left(1- \frac{m_\ell^2}{m_N^2} \right)^2\nonumber\\
 &  & + \frac{m_X^2}{m_N^2} \left(1+ \frac{m_\ell^2}{m_N^2} \right) - 2 \left(\frac{m_X^2}{m_N^2}\right)^2 \Bigg ], \nonumber\\
\label{dwidth22}
\end{eqnarray}
where $f_X$ is the hadron decay constant, see Table \ref{table2}. 

\begin{table}[h]
\caption{\label{table2} Mass and decay constant mesons.}
\begin{ruledtabular}
\begin{tabular}{ccc}
\textrm{Particle} & \textrm{Mass[MeV]} &  $f_X$ \textrm{[MeV]} \\
\colrule
$\pi^\pm$ & 139.57 & 130.41 \\
$K^\pm$ & 493.67 & 156.2 \\
$\rho^\pm$ & 775.49 & 220 \\
$K^{*\pm}$ & 775.49 & 217 \\
\end{tabular}
\end{ruledtabular}
\end{table}

The lifetime of the Majorana neutrino $\tau_N = \hslash /\Gamma_N$ in Eq. (\ref{eq:factorForm}) can be obtained by summing over all accessible final states that can be opened at the mass $m_N$ \citep{Abe:2010gxa}. However, in further analysis (Secs. \ref{sec3} and \ref{sec4}), we will leave it as a phenomenological parameter accessible to the Belle and Belle II experiments. 

\section{\label{sec3}Expected experimental sensitivity at Belle and Belle II}

Now, let us provide an estimation of the expected number of events at the SuperKEKB  \citep{Abe:2010gxa}, namely Belle II experiment and its predecessor Belle \citep{Bevan2014}, for the $|\Delta L| = 2$ channels of the $\tau$ lepton, $\tau^- \to X^+ \ell^- \ell^- \nu_\tau$ (with $X = \pi, K, \rho, K^*$), discussed above.

\subsection{\label{sec3.1}Belle and Belle II experiments}

The SuperKEKB accelerator is upgraded from KEKB \citep{Bevan2014} and its target luminosity is $8\times 10^{35}$ cm$^{-2}$s$^{-1}$,  40 times higher than KEKB. SuperKEKB collides electrons and positrons at the $\Upsilon(4S)$ resonance energy, producing a large amount of B meson pairs, a B-factory. However, the cross section of the process $e^+e^- \to \tau^+\tau^-$ at the $\Upsilon(4S)$ resonance energy is of the same order as the production of a B pair, then, SuperKEKB is also a $\tau$ lepton factory.

Furthermore, the four-body $|\Delta L| = 2$ channel of $\tau^- \to X^+ \ell^- \ell^- \nu_\tau$ decays haven't been studied neither Belle and Belle II. Then, let us provide an estimation of the expected number of events at these experiments, which has the form  

\begin{eqnarray}
 N_{\rm exp}^{\rm Belle/Belle\,II} &=&  \sigma(ee \to \tau\tau){\rm BR}(\tau^{\pm} \to \Delta L=2)\nonumber \\
  && \times  \epsilon_D^{\rm Belle/Belle II}(\tau^{\pm} \to \Delta L=2)\nonumber\\
  && \times \mathcal{L}^{\rm Belle/Belle\,II}_{\rm int},
 \label{N:belle}
\end{eqnarray}
where $\sigma(ee \to \tau\tau)$ is the production cross section of $\tau$ pairs inside the Belle/Belle II geometrical acceptance; $\mathcal{L}^{\rm Belle/Belle\,II}_{\rm int}$ is the integrated luminosity; BR$(\tau^{\pm} \to \Delta L=2)$ correspond to the branching fraction of the given LNV process and $\epsilon_D^{\rm Belle/BelleII}(\tau^{\pm} \to \Delta L=2)$ is its detection efficiency of the Belle/Belle II detector involving reconstruction, selection, trigger, particle misidentification and detection efficiencies. 

The production cross section has been measured to $\sigma(ee \to \tau\tau) = 0.919\pm0.003$ nb \citep{Bevan2014} inside the Belle acceptance and for Belle II is expected the same behavior \citep{Kou:2018nap}. The proper computation of the detection efficiency requires fully simulated Monte Carlo samples of the exclusive decay, reconstructed in the same way as real Belle/Belle II data. Here, we perform a rough estimation of the detection efficiency, based on extrapolation of detection efficiencies already reported by Belle experiment of similar final states.

The Belle Collaboration has measured the detection efficiency of $\tau$ decay modes to be $2.73\pm0.10\%$ for $\tau^- \to \pi^- e^+ e^- \nu_\tau$ and $4.14\pm0.16\%$ for $\tau^- \to \pi^- \mu^+ \mu^- \nu_\tau$ \citep{PhysRevD.100.071101}. This measurement includes trigger, tracking, reconstruction, particle identification, and selection efficiency. Given the content of final-state charged tracks, we can consider the $\tau^- \to \pi^+ e^- e^- \nu_\tau$ detection efficiency to be the same as for the $\tau^- \to \pi^- e^+ e^- \nu_\tau$ decays. In the case of $\tau^- \to \pi^+ \mu^- \mu^- \nu_\tau$ detection efficiency we consider the same as $\tau^- \to \pi^- \mu^+ \mu^- \nu_\tau$ decays.
On the other hand, the reconstruction efficiency is approximately $10\%$ lower for kaon than for pion \citep{MIYAZAKI2013346, PhysRevD.88.052019}. Thus, we can multiply the previous efficiencies $2.73\pm0.10\%$ and $4.14\pm0.16\%$ by 0.9 for the cases $\tau^- \to K^+ e^- e^- \nu_\tau$ and $\tau^- \to K^+ \mu^- \mu^- \nu_\tau$, respectively. 
The $\rho^+$ reconstruction implies an extra reconstruction of $\pi^0$ with a systematic detection error of $3 \%$ \citep{PhysRevD.76.011104, PhysRevLett.94.121801,CHANG2004148}. The $K^{*+}$ reconstruction efficiency is approximately $95\%$ \citep{PhysRevD.88.052019}. 
In general, we consider efficiencies of $2.457\pm0.20\%$ and $3.726\pm0.20\%$ for $\tau^- \to X^+ e^- e^- \nu_\tau$ and $\tau^- \to X^+ \mu^- \mu^- \nu_\tau$, respectively.

\begin{figure}[h]
\begin{center}
  \includegraphics[width=1.05\linewidth]{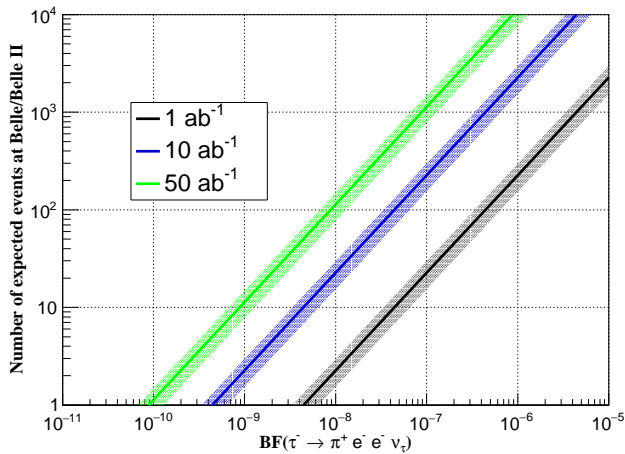}
\end{center}
\caption{\label{fig:eventEE} Number of expected events at Belle (1 ab$^{-1}$) and Belle II (50 ab$^{-1}$) for $\tau^- \to \pi^+ e^- e^- \nu_\tau$ as a function of the branching fraction, for different luminosity values: 1 ab$^{-1}$ (black), 10 ab$^{-1}$ (blue) and 50 ab$^{-1}$ (green). The filled region represent the 1-$\sigma$ uncertainty.}
\end{figure}

\begin{figure}[h]
\begin{center}
  \includegraphics[width=1.05\linewidth]{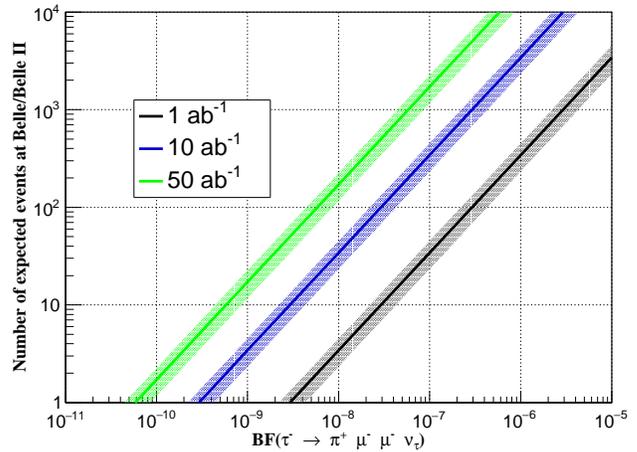}
\end{center}
\caption{\label{fig:eventMuMu}  Number of expected events at Belle (1 ab$^{-1}$) and Belle II (50 ab$^{-1}$) for $\tau^- \to \pi^+ \mu^- \mu^- \nu_\tau$ as a function of the branching fraction, for different luminosity values: 1 ab$^{-1}$ (black), 10 ab$^{-1}$ (blue) and 50 ab$^{-1}$ (green). The filled region represent the 1-$\sigma$ uncertainty..}
\end{figure}

Also, we have considered three different scenarios, $\mathcal{L} = 1$, 10 and 50 ab$^{-1}$, which approximately correspond to the Belle sample collected at the KEKB nominal construction energy of a center of mass of 10.54 GeV. On the other hand, Belle II expect to collect an integrated luminosity of 50 ab$^{-1}$. Assuming the above assumptions on efficiency and cross section, figures  \ref{fig:eventEE} and \ref{fig:eventMuMu} shows the number of expected events to be observed in Belle and Belle II experiments as a function of branching fraction for $|\Delta L| = 2$ four-body modes of $\tau$ lepton. The figure shows black, blue and green functions, corresponding to Belle (1 ab$^{-1}$) and Belle II (10 and 50 ab$^{-1}$), respectively. We found a significant sensitivity at the Belle on branching fraction of the order $10^{-9} - 10^{-8}$ for $\tau^- \to X^+ \ell^- \ell^- \nu_\tau$. Furthermore,  this sensitivity increase in Belle II reaching a branching fraction of the order $10^{-10} - 10^{-9}$.

In the analysis of the next section, we will take a branching fraction of the order of $10^{-9}$ as the conservative and accessible to Belle II and the limit for Belle.  

\section{\label{sec4}Bounds on the parameter space $(m_N, |V_{\ell N}|^2)$}

The experimental non-observation of $|\Delta L| = 2$ processes can be reinterpreted as bounds on the parameter space of a heavy sterile neutrino $(m_N, |V_{\ell N}|^2)$, namely, the squared mixing element $|V_{\ell N}|^2$ as function of the mass $m_N$. Based on the analysis presented in Sec. \ref{sec3}, here, we explore the constraints on the $(m_N, |V_{\ell N}|^2)$ plane that can be achieved from the experimental searches on $\tau^- \to X^+ \ell^- \ell^- \nu_\tau$ at the SuperKEKB, namely the Belle II experiment as well as its predecessor Belle. 

From Eqs. (\ref{eq:factorForm}), (\ref{eq:wideeq4}) and (\ref{dwidth21}) or (\ref{dwidth22}) , it is straightforward to obtain the relation 

\begin{equation}
  |V_{\ell N}|^2 = \left[ \frac{BR(\tau^- \to X^+ \ell^- \ell^- \nu_\tau)  \hslash}{\overline{BR}(\tau^- \to \nu_\tau \ell^- N)\times \overline{\Gamma}(N \to \ell^- X^+) \tau_N} \right]^{1/2}
\end{equation}
with $\overline{BR}(\tau^- \to \nu_\tau \ell^- N) = BR(\tau^- \to \nu_\tau \ell^- N)/|V_{\ell N}|^2$ and $\overline{\Gamma}(N \to \ell^- X^+) = \Gamma(N \to \ell^- X^+)/|V_{\ell N}|^2$. We will consider the heavy neutrino lifetime of $\tau_N$ = 5, 100 ps, which corresponds to an average flight distance of up to 30 mm, well inside the Belle vertex detector \cite{Bevan2014}. This will allows us to extract the limits on $|V_{\ell N}|^2$ without any additional assumption on the relative size of the mixing matrix elements.

\begin{figure}[h]
\begin{center}
  \includegraphics[width=1\linewidth]{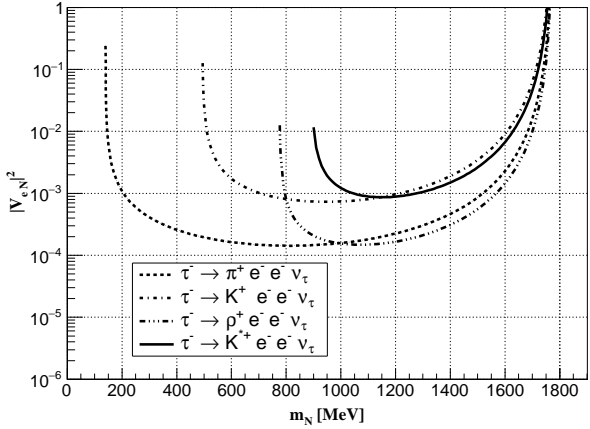}
\end{center}
\caption{\label{fig:regionEE_5ps} Exclusion region on the ($m_N, |V_{eN}|^2$) plane for BR($\tau^- \to X^+ e^- e^- \nu_\tau) < 10^{-9}$   (with $X = \pi$, $K$, $\rho$ or $K^*$). These regions represent the bounds obtained for heavy neutrino lifetime of $\tau_N = 5$ ps.}
\end{figure}

\begin{figure}[h]
\begin{center}
  \includegraphics[width=1\linewidth]{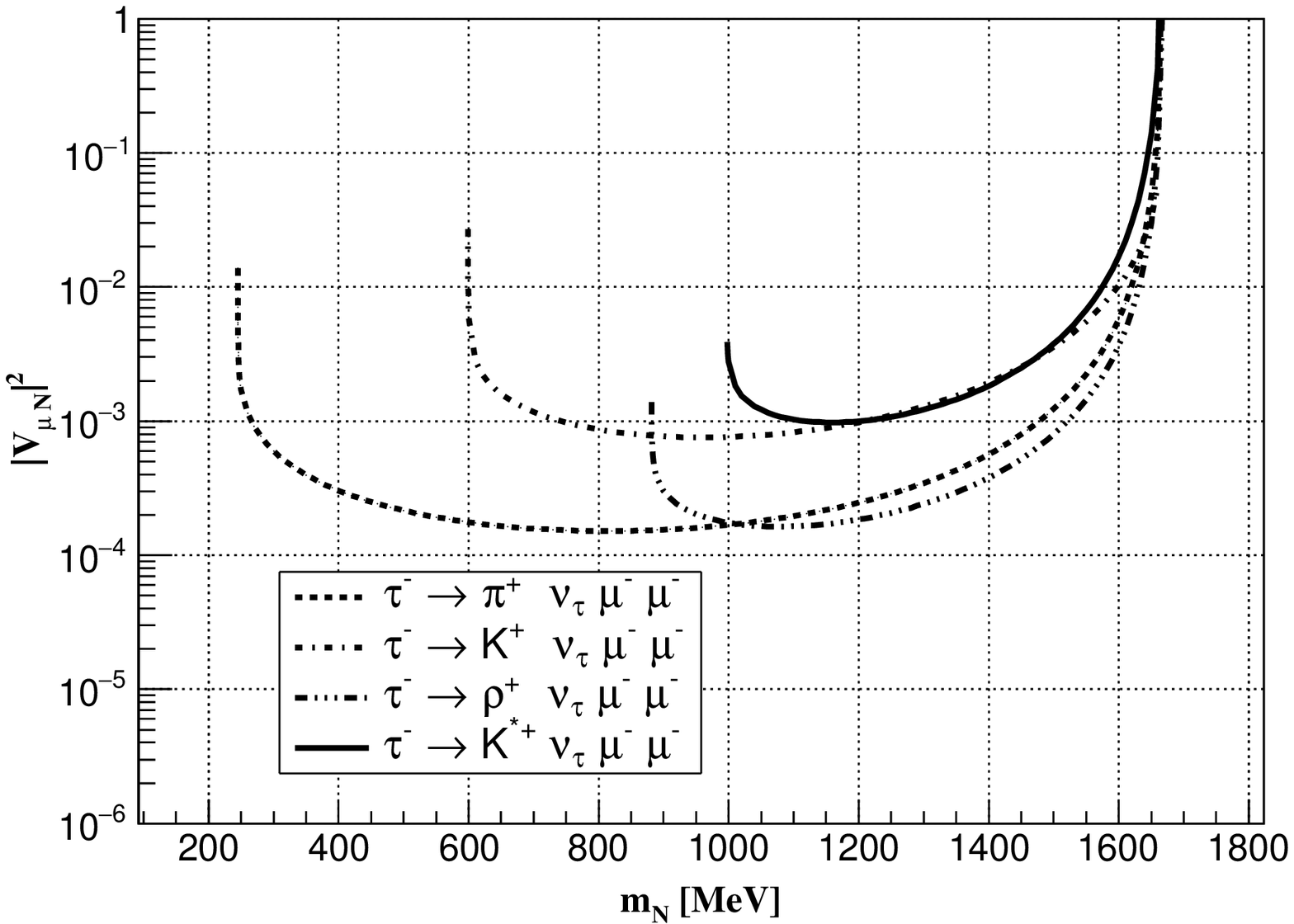}
\end{center}
\caption{\label{fig:regionMuMu_5ps} Exclusion region on the ($m_N, |V_{\mu N}|^2$) plane for BR($\tau^- \to X^+ \mu^- \mu^- \nu_\tau) < 10^{-9}$   (with $X = \pi$, $K$, $\rho$ or $K^*$). These regions represent the bounds obtained for heavy neutrino lifetime of $\tau_N = 5$ ps.}
\end{figure}

\begin{figure}[h]
\begin{center}
  \includegraphics[width=1\linewidth]{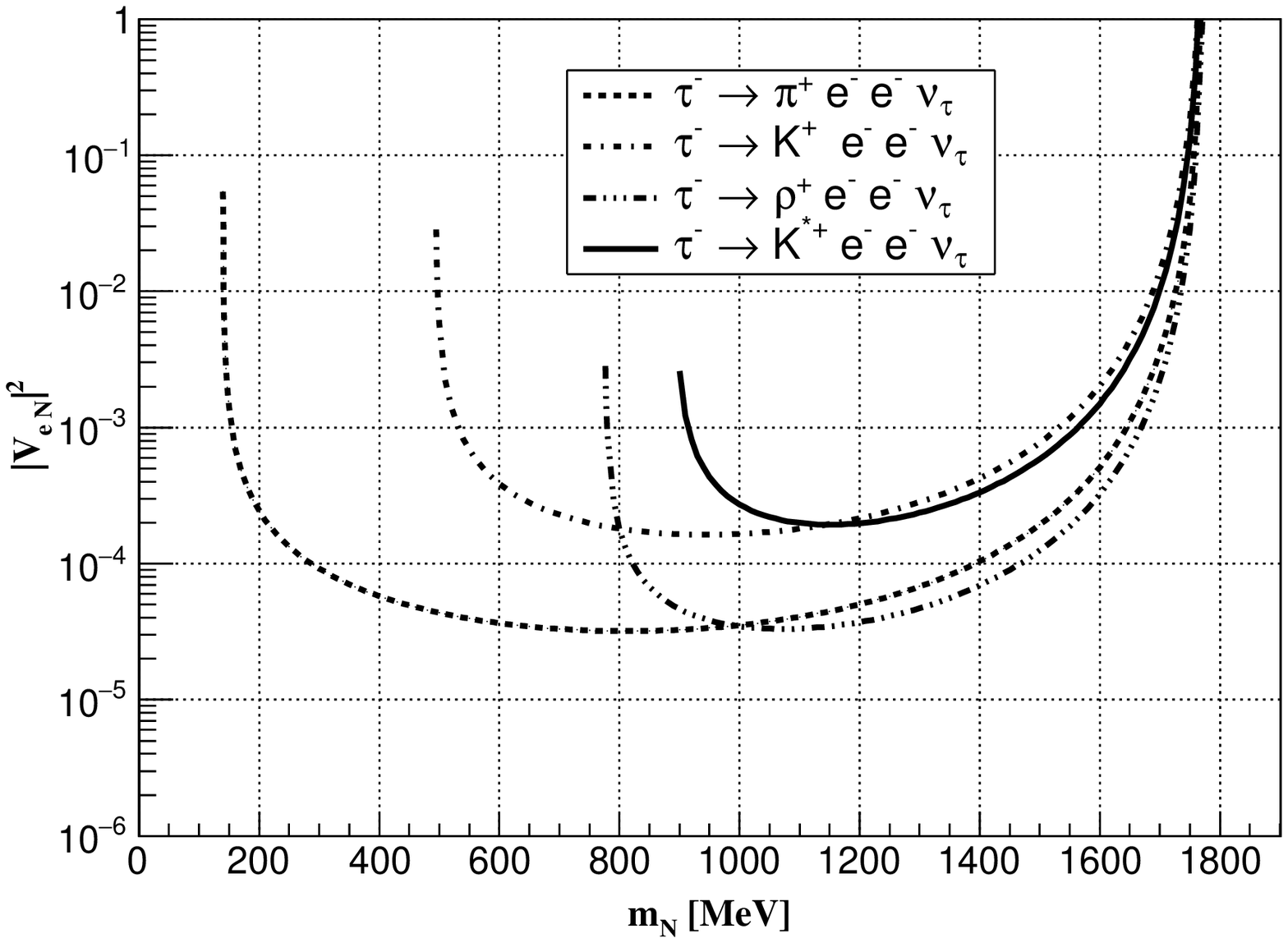}
\end{center}
\caption{\label{fig:regionEE_100ps} Exclusion region on the ($m_N, |V_{eN}|^2$) plane for BR($\tau^- \to X^+ e^- e^- \nu_\tau) < 10^{-9}$   (with $X = \pi$, $K$, $\rho$ or $K^*$). These regions represent the bounds obtained for heavy neutrino lifetime of $\tau_N = 100$ ps.}
\end{figure}

\begin{figure}[h]
\begin{center}
  \includegraphics[width=1\linewidth]{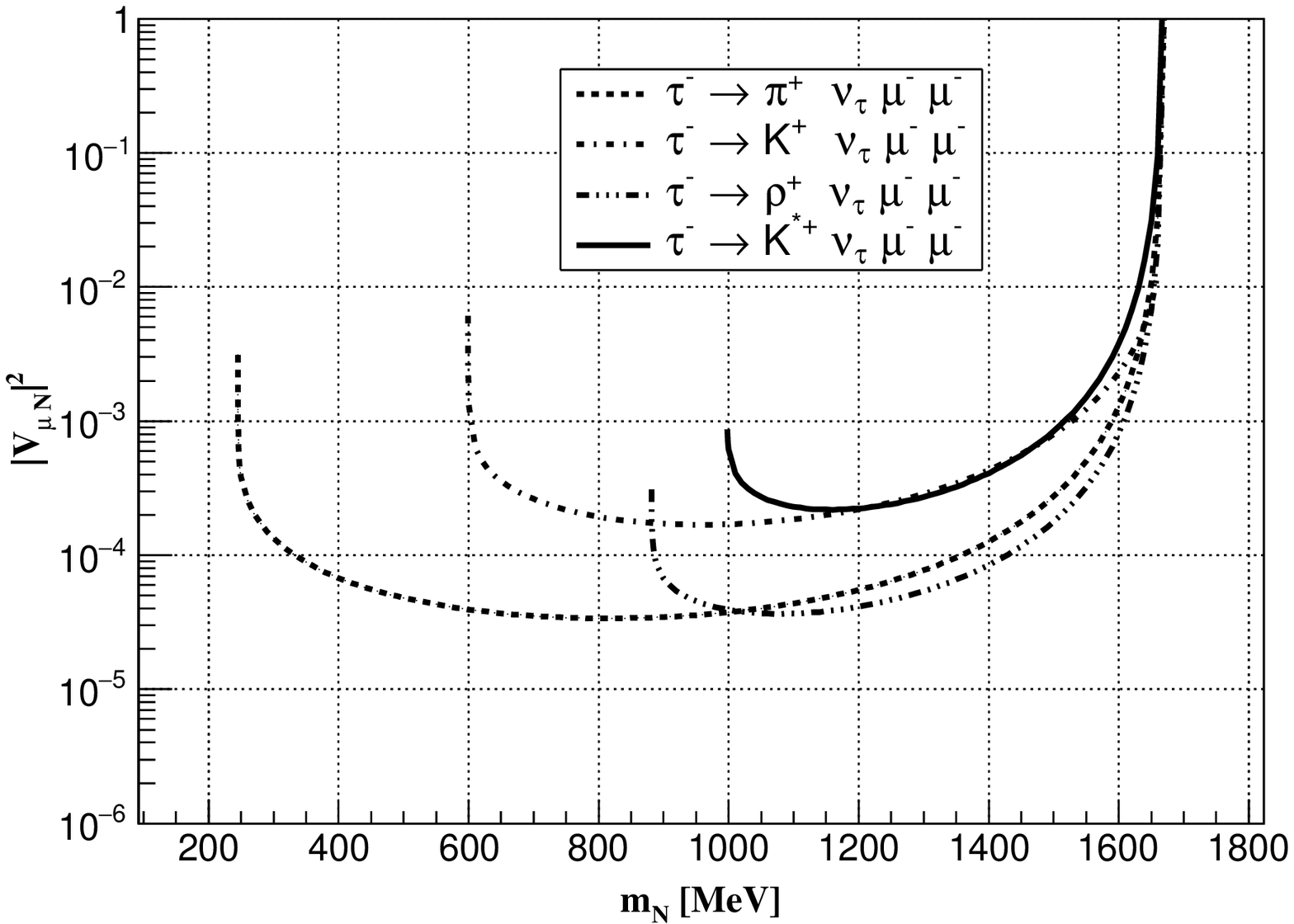}
\end{center}
\caption{\label{fig:regionMuMu_100ps} Exclusion region on the ($m_N, |V_{\mu N}|^2$) plane for BR($\tau^- \to X^+ \mu^- \mu^- \nu_\tau) < 10^{-9}$   (with $X = \pi$, $K$, $\rho$ or $K^*$). These regions represent the bounds obtained for heavy neutrino lifetime of $\tau_N = 100$ ps.}
\end{figure}

Considering an expected Belle II sensitivity on the branching fractions of the order $BR(\tau^- \to X^+ \ell^- \ell^- \nu_\tau) < 10^{-9}$ for 10 ab$^{-1}$, in Figs. (\ref{fig:regionEE_5ps}) and (\ref{fig:regionEE_100ps}) we show the exclusion regions on the $(m_N, |V_{e N}|^2)$ plane obtained from future searches on $\tau^- \to X^+ e^- e^- \nu_\tau$ at 5 and 100 ps, respectively. In Figs. (\ref{fig:regionMuMu_5ps}) and (\ref{fig:regionMuMu_100ps}) we show the exclusion regions for the corresponding $(m_N, |V_{\mu N}|^2)$ on $\tau^- \to X^+ \mu^- \mu^- \nu_\tau$ at 5 and 100 ps, respectively.

In Figs. (\ref{fig:regionEE_5ps}) and (\ref{fig:regionMuMu_5ps}),  we can observe that the most restrictive bound is given by $\tau^- \to \pi^+ e^- e^- \nu_\tau$ and $\tau^- \to \pi^+ \mu^- \mu^- \nu_\tau$, respectively. Which can reach $|V_{\ell N}|^2 \sim \mathcal{O}(10^{-4})$ at $\tau_N = 5$ ps and $|V_{\ell N}|^2 \sim \mathcal{O}(10^{-5})$ at $\tau_N = 100$ ps for a large mass window of  [0.140, 1.776] GeV for $\tau^- \to \pi^+ e^- e^- \nu_\tau$ and [0.245, 1.671] GeV for $\tau^- \to \pi^+ \mu^- \mu^- \nu_\tau$.

\section{\label{sec5}Conclusions}

We have explored a $\tau$ search to track the possible signals of lepton-number-violation at the Belle and Belle II experiments, due to the considerably tau pair production. We studied the four-body $|\Delta L| = 2$ decays of the $\tau$ lepton, $\tau^- \to X^+ \ell^- \ell^- \nu_\tau$ ($X = \pi$, $K$, $\rho$, $K^*$), induced by an on-shell Majorana neutrino $N$ with a mass of few GeV and lifetime of $\tau_N$ = 5, 100 ps, well inside the Belle vertex detector. This to extract the limits on $|V_{\ell N}|^2$ without any additional assumption on the relative size of the mixing matrix elements. We performed an exploratory study on the potential sensitivity that Belle II experiment that could achieve for these $|\Delta L| = 2$ processes as well as the limit for Belle experiment. For a long term expected integrated luminosity at Belle II of 10 ab$^{-1}$, we found that branching fractions of the order $\mathcal{O}(10^{-9})$ might be feasible. Such sensitivity will allow to cover a neutrino mass window of $0.140 < m_N < 1.776$ GeV for $\tau^- \to X^+ e^- e^- \nu_\tau$ and $0.245 < m_N < 1.671 $ GeV for $\tau^- \to X^+ \mu^- \mu^- \nu_\tau$ corresponding to a squared mixing element $|V_{\ell N}|^2 \sim \mathcal{O}(10^{-5})$ at $\tau_N = 100$ ps.

\begin{acknowledgments}
Support for this work has been received in part by Consejo Nacional de Ciencia y Tecnolog\'ia grant number A1-S-33202. 
\end{acknowledgments}

\bibliography{version1}

\end{document}